\documentclass[aip, pop, amsmath,amssymb, preprint,floatfix]{revtex4-1}

\usepackage{color}
\usepackage[table]{xcolor} 
\usepackage{graphicx}
\usepackage{dcolumn}
\usepackage{bm}
\usepackage{psfrag}
\usepackage{epstopdf}
\usepackage{subcaption}
\usepackage{textcomp}
\usepackage{hyperref}
\usepackage{comment}
\usepackage{multirow}
\usepackage{amstext} 

\usepackage{sidecap}

\newcommand{\nebar}{$<$$n_e$$>$}

\newcommand{\Ip}{$I$}
\newcommand{\Bt}{$B$}
\newcommand{\Rnot}{$R_0$}
\newcommand{\aminor}{$a$}
\newcommand{\betat}{$\beta_{T}$}
\newcommand{\betan}{$\beta_{N}$}
\newcommand{\qnf}{$q_{95}$}
\newcommand{\elong}{$\kappa$}
\newcommand{\triavg}{$\delta_{avg}$}
\newcommand{\PoverPLH}{$P_{net}/P_{LH08}$}

\newcommand{\Ptot}{$P_{tot}$}
\newcommand{\Pech}{$P_{ECH}$}
\newcommand{\Pnet}{$P_{net}$}
\newcommand{\PecoverPtot}{$P_{ECH}/P_{tot}$}

\newcommand{\Paux}{$P_{aux}$}

\newcommand{\IaB}{$IaB$}
\newcommand{\In}{$I/aB$}
\newcommand{\lawson}{$\left<p \right>$$\tau$}
\newcommand{\snyder}{$\left<p \right>$$\tau/IaB$}
\newcommand{\HH}{$H_{H98y2}$}
\newcommand{\HL}{$H_{L89}$}

\newcommand{\pres}{$\left<p \right>$}
\newcommand{\rhostar}{$\rho_i^*$}
\newcommand{\nustar}{$\nu_e^*$}
\newcommand{\taue}{$\tau_E$}
\newcommand{\taueth}{$\tau_{E,th}$}

\newcommand{\pped}{$p_{ped}$}
\newcommand{\peped}{$p_{e,ped}$}

\newcommand{\navg}{$\left<n_e \right>$}
\newcommand{\neped}{$n_{e,ped}$}
\newcommand{\teped}{$T_{e,ped}$}

\newcommand{\density}{$\left<n_e \right>$}

\newcommand{\supr}{RMP-ELM suppression}
\newcommand{\supred}{RMP-ELM suppressed}

\newcommand{\diiid}{{\color{black}{DIII-D}}}
\newcommand{\aug}{{\color{black}{AUG}}}
\newcommand{\east}{{\color{black}{EAST}}}
\newcommand{\kstar}{{\color{black}{KSTAR}}}
\newcommand{\devices}{\aug{}, \diiid{}, \east{}, and \kstar{}}

\begin{document}

\preprint{}

\title[]{Plasma Performance and Operational Space with an RMP-ELM Suppressed Edge}

\author{C. Paz-Soldan}
\affiliation{Dept. of Applied Physics \& Applied Mathematics, Columbia University, New York, NY, USA}
\email{carlos.pazsoldan@columbia.edu}

\author{S. Gu}
\affiliation{Oak Ridge Associated Universities, Oak Ridge, TN, USA}
\affiliation{Institute of Plasma Physics, Chinese Academy of Sciences, Hefei, China}

\author{N. Leuthold}
\affiliation{Dept. of Applied Physics \& Applied Mathematics, Columbia University, New York, NY, USA}
\affiliation{Oak Ridge Associated Universities, Oak Ridge, TN, USA}

\author{P. Lunia}
\affiliation{Dept. of Applied Physics \& Applied Mathematics, Columbia University, New York, NY, USA}

\author{P. Xie}
\affiliation{Institute of Plasma Physics, Chinese Academy of Sciences, Hefei, China}

\author{M.W. Kim}
\affiliation{Korea Institute of Fusion Energy, Daejon, Republic of Korea}

\author{S.K. Kim}
\affiliation{Princeton Plasma Physics Laboratory, Princeton, NJ USA}

\author{N.C. Logan}
\affiliation{Dept. of Applied Physics \& Applied Mathematics, Columbia University, New York, NY, USA}

\author{J.-K. Park}
\affiliation{Department of Nuclear Engineering, Seoul National University, Seoul, Republic of Korea}

\author{W. Suttrop}
\affiliation{Max-Planck-Institut fur Plasmaphysik, EURATOM Association, Garching, Germany}

\author{Y. Sun}
\affiliation{Institute of Plasma Physics, Chinese Academy of Sciences, Hefei, China}

\author{D.B. Weisberg}
\affiliation{General Atomics, San Diego, CA 92121, USA}

\author{M. Willensdorfer}
\affiliation{Max-Planck-Institut fur Plasmaphysik, EURATOM Association, Garching, Germany}

\author{the ASDEX-Upgrade}
\affiliation{See Stroth et al. 2022 (\url{https://doi.org/10.1088/1741-4326/ac207f}) for the ASDEX Upgrade Team.}

\author{DIII-D}

\author{EAST}

\author{KSTAR Teams}



\vspace{-0.28 in}
\begin{abstract}

The operational space and global performance of plasmas with edge-localized modes (ELMs) suppressed by resonant magnetic perturbations (RMPs) are surveyed by comparing \devices{} stationary operating points. \supr{} is achieved over a range of plasma currents, toroidal fields, and RMP toroidal mode numbers. Consistent operational windows in edge safety factor are found across devices, while windows in plasma shaping parameters are distinct. Accessed pedestal parameters reveal a quantitatively similar pedestal-top density limit for RMP-ELM suppression in all devices of just over $3\times 10^{19}$ m$^{-3}$. This is surprising given the wide variance of many engineering parameters and edge collisionalities, and poses a challenge to extrapolation of the regime. Wide ranges in input power, confinement time, and stored energy are observed, with the achieved triple product found to scale like the product of current, field, and radius. Observed energy confinement scaling with engineering parameters for \supred{} plasmas are presented and compared with expectations from established H and L-mode scalings, including treatment of uncertainty analysis. Different scaling exponents for individual engineering parameters are found as compared to the established scalings. However, extrapolation to next-step tokamaks ITER and SPARC find overall consistency within uncertainties with the established scalings, finding no obvious performance penalty when extrapolating from the assembled multi-device \supred{} database. Overall this work identifies common physics for \supr{} and highlights the need to pursue this no-ELM regime at higher magnetic field and different plasma physical size.

\end{abstract}


\maketitle



\section{Introduction and Motivation}
\label{sec:intro}

Extrapolation of the tokamak approach to fusion energy production is challenged by the repetitive Edge Localized Mode (ELM) instability \cite{Zohm1996,Leonard2014}. This instability is driven by edge gradients, and at reactor scale delivers a damaging heat and particle load to the first-wall\cite{Loarte2003a,Federici2003,Dux2011,Pitts2013}. The ELM also poses a risk to stable plasma operation due to high-Z wall material ingress into the plasma \cite{Kallenbach2005,Beurskens2014,Kim2018}. For this reason, ELM control is considered essential to operate at sufficient plasma current to achieve the ITER Q=10 mission \cite{Loarte2014,Maingi2014}.

ELM suppression by resonant magnetic perturbations (RMPs) is a primary technique to control the ELM in ITER, and the physics of \supr{} has recently been thoroughly reviewed in Ref. \cite{ITPAPEPNF2023} and before that in Ref. \cite{Evans2015}. With the RMP technique, application of $\approx$0.1\% level non-axisymmetric fields from nearby window-pane coils maintains a fully formed pedestal yet prevents the ELM - but only if certain access criteria are met. When \supr{} is achieved, enhanced transport is observed, arising from either: parallel transport across macroscopic large-scale pedestal-top magnetic islands \cite{Waelbroeck2009,Nazikian2015,Hu2019,Fitzpatrick2020a}; increased turbulent fluctuation levels directly arising from field penetration \cite{McKee2013,Lee2016,Lee2019}; increased turbulence arising indirectly from penetration-induced changes to the radial electric field structure \cite{Sung2017b,Taimourzadeh2019}; or a combination of these effects. Additional mechanisms for \supr{} have also been proposed, such as neoclassical transport effects \cite{Callen2012b,Huijsmans2015} or localized peeling-ballooning instabilities driven by 3D equilibrium modifications \cite{Bird2013,Willensdorfer2017,Ryan2019,Kim2020a}. These mechanisms do not require significant field penetration, and as such struggle to explain the pedestal bifurcation from RMP-ELM mitigation into full ELM suppression \cite{Nazikian2015}.

\begin{figure*}
\centering
\includegraphics[width=1\textwidth]{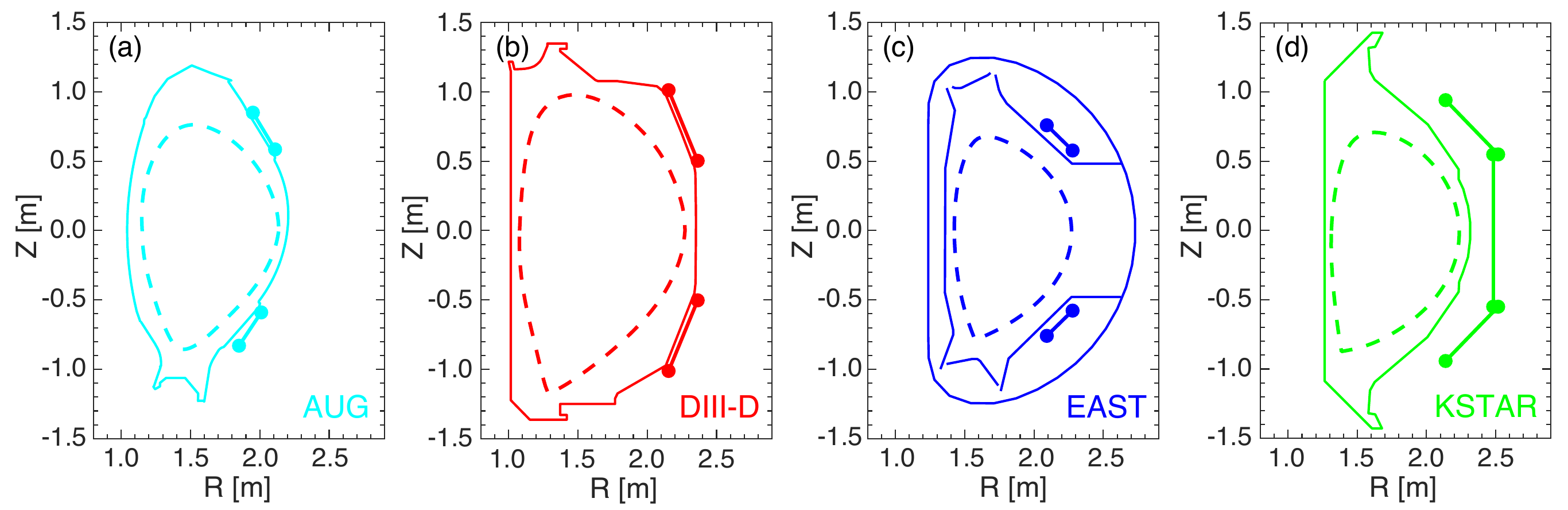} 
 \vspace{-20 pt}
\caption{Comparison of vacuum vessel, RMP coil, and typical plasma geometry of the mid-scale tokamak devices that are the focus of this study: (a) \aug{}, (b) \diiid{}, (c) \east{}, (d) \kstar{}.}
\label{fig:devices}
\end{figure*}

After its initial discovery in \diiid{} in 2003 \cite{Evans2004,Moyer2005,Burrell2005,Evans2006}, \supr{} has been exported to tokamaks worldwide, and RMP coilsets have been incorporated into the ITER design \cite{Schaffer2008,Becoulet2008}. RMP coils are also planned for ELM control in SPARC \cite{Sweeney2020}. In \kstar{}, RMP coils with a unique engineering design were installed in 2008 \cite{Kim2009}, H-mode operation was established in 2010, and \supr{} accessed in 2011 \cite{Jeon2012}. In \aug{}, the RMP coil system was installed in two stages between 2010-2011 \cite{Suttrop2009}. Following improvements in scenario development to access low-collisionality pedestal regimes together with an increase in the plasma triangularity, full ELM suppression on \aug{} was accessed in 2016 \cite{Nazikian2016IAEA,Suttrop2018}. In \east{}, RMP coils were installed in 2014 \cite{Sun2015} and \supr{} was accessed in dedicated experiments in 2015 \cite{Sun2016}. The geometry of these devices, including their RMP coils, is presented in Fig.~\ref{fig:devices}. Resulting from concerted effort over many years, the access to \supr{} in different tokamaks worldwide offers a unique opportunity to explore cross-machine comparison and projection of the RMP technique, which this study begins to undertake. 

It should also be noted that RMP coils have been installed in devices achieving plasma currents in the mega-Amp range without yielding access to \supr{}. Databases for these devices with RMP coils active but with persistent ELMs have not been compiled and are beyond the scope of this work. Briefly, however, it can be said that these devices fall into two categories. First, devices with distant RMP coils that couple poorly to the edge resonant surfaces, such as JET \cite{Barlow2001,Liang2007} and Alcator C-mod \cite{Wolfe2005}; and second, the spherical tokamak devices MAST(-U) \cite{Kirk2012} and NSTX(-U) \cite{Canik2010}, for which the absence of \supr{} is not yet understood. \supr{} has also not been achieved in devices with currents well below a mega-Amp, presumably because sufficiently collisionless conditions cannot be achieved at that scale.
 
A first goal of this work is to document and compare the operational space accessed by tokamak plasmas with \supred{} edges. It is known that access to this regime requires specific operational criteria. However, after dedicated experimentation,  the regime has been found on several tokamak devices worldwide. Each device exploring \supr{} has unique characteristics, thus comparing the operational space can reveal areas of universality for \supr{}, as well as areas of discrepancy. Furthermore, since many engineering parameters vary significantly across these devices, exploration of the plasma performance in terms of confinement and triple product \cite{Lawson1957} metrics can reveal empirical trends. Exploring these trends is a second goal of this work, alongside performing engineering regressions of the global confinement time to compare to trends in the established H- and L-mode scalings. 

While this work focuses on \supr{}, this is only one technique among several being explored to achieve ELM control in ITER, SPARC, and future reactors. These other approaches (such as Quiescent H-mode, Improved confinement I-mode, Enhanced D-alpha mode, and small/grassy ELM regimes) have been well-reviewed in the literature \cite{Oyama2006,Viezzer2018}. A comparison of the operating space and plasma performance in DIII-D of the various no-ELM regimes has also been published recently \cite{PazSoldan2021}. Additionally, a similarly motivated multi-device comparison of I-mode can be found in the literature \cite{Hubbard2016}.



Criteria for inclusion of a datapoint into this database are largely the same as was elaborated in Ref. \cite{PazSoldan2021}, where a detailed discussion of each criteria is included. A key criteria is stationarity, judged manually via the identification of stationary phases where parameters are roughly constant. The stationarity filter results in a subset of \supred{} points being retained. Additionally, machine parameters (field, current, power) must be fixed, ELMs must not be present during the time window (ie, full \supr{}), and the time windows must be longer than approximately three energy confinement times. Finally, all data presented is from deuterium main-ion plasmas.

Furthermore, this work has endeavored to be as comprehensive as feasible. For \aug{} and \east{}, new databases were populated by examining every \supred{} discharge since the discovery of the regime on those devices (2015 on \east{}, 2016 on \aug{}) until 2020. This resulted in $\approx$ 200 stationary phases for \aug{}, and $\approx$ 250 stationary phases for \east{}. For \kstar{}, the database first presented in Ref. \cite{Kim2020c} was reused for this work, and the equilibria were recomputed using improved sensor compensation techniques. This \kstar{} database features a comprehensive survey of ELM-suppressed discharges from 2016-2019. For \diiid{}, the database first presented in Ref. \cite{PazSoldan2021} is reused. Owing to the maturity of the RMP technique in \diiid{}, it is impractical to survey every \supred{} discharge, of which several thousand qualify. Instead, a curated subset of discharges highlighting variations in operating space and plasma performance is presented. Additionally, in some figures, data from a global database of type-I ELMing discharges in H-mode will be shown. This `DB4v5' database was prepared by the International Tokamak Physics Activity (ITPA) for the improvement of the standard \HH{} confinement scaling law \cite{Thomsen2002}. 


The organization of this paper is divided into an exposition of the \supred{} operating space in Sec.~\ref{sec:ops} and a discussion of observed plasma performance and confinement scaling in Sec.~\ref{sec:perf}. The operating space is presented in terms of basic machine parameters (Sec.~\ref{sec:eng}), RMP coil settings (Sec.~\ref{sec:RMP}), pedestal parameters (Sec.~\ref{sec:corr}), and finally electron heating (Sec.~\ref{sec:ech}). The plasma performance is discussed in terms of normalized performance (Sec.~\ref{sec:norm}), absolute performance (Sec.~\ref{sec:abs}), and confinement time scaling (Sec.~\ref{sec:scaling}). Discussions and conclusions are given in Sec.~\ref{sec:disc}.


\section{RMP-ELM Suppressed Operating Space}
\label{sec:ops}

Observations in terms of machine parameters, RMP coil settings, pedestal parameters, and electron heating response are described sequentially to document the plasma operating space with an \supred{} edge. Since each datapoint is a stationary window with an \supred{} edge, the presented figures can be interpreted as windows of \supr{} amidst a background of ELMing datapoints that are generally not shown. In some plots, the operating space is defined by the return of the ELM, while in others, it represents a more fundamental limitation of device capability. Note discussion of divertor integration, while found in Ref \cite{PazSoldan2021} for \diiid{}, is outside of the scope of this work. This is because divertor studies are idiosyncratic due to the variable divertor geometries found across devices. In contrast, the parameters highlighted here are expected to be more generically representative of the \supred{} edge.

\begin{figure*}
\begin{subfigure}{1\textwidth}
\begin{subfigure}{0.32\textwidth}
\centering
\includegraphics[width=1\textwidth]{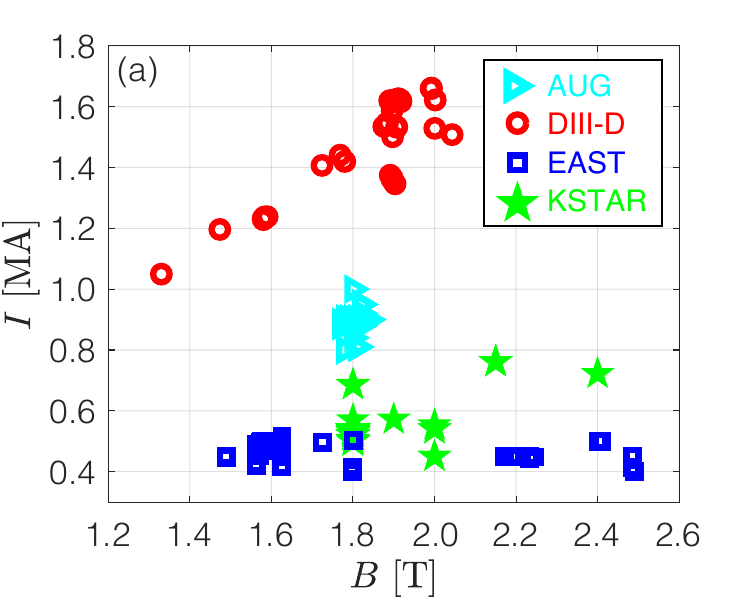}
\end{subfigure}
\begin{subfigure}{0.34\textwidth}
\centering
\includegraphics[width=1\textwidth]{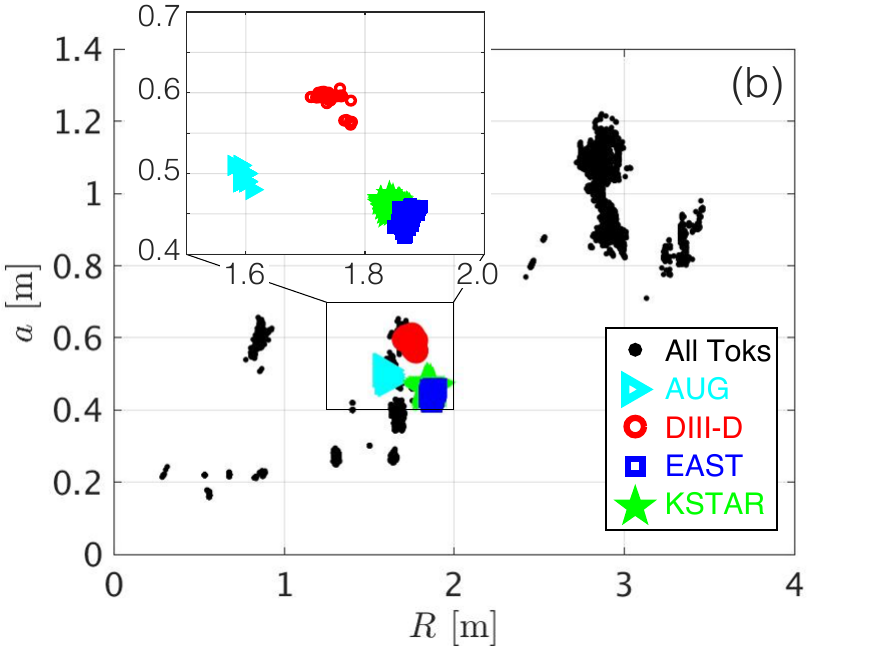} 
\end{subfigure}
\begin{subfigure}{0.32\textwidth}
\centering
\includegraphics[width=1\textwidth]{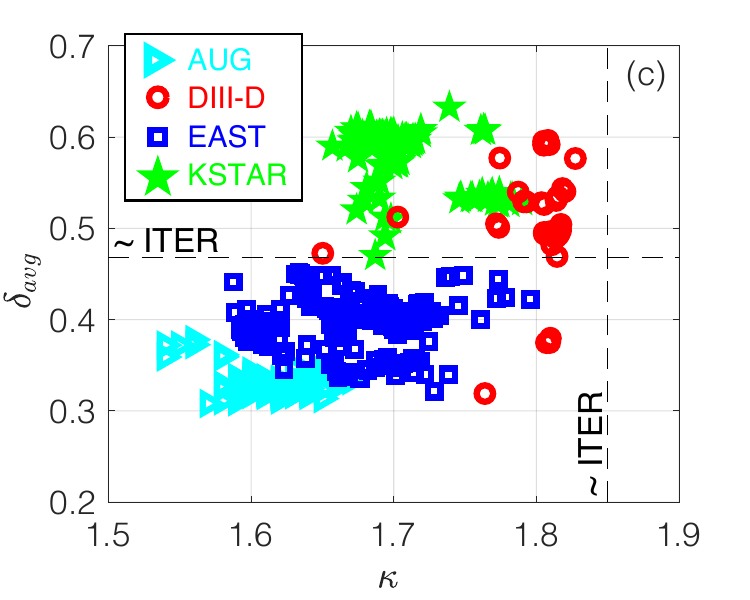}
\end{subfigure}
\end{subfigure}
 \vspace{-5 pt}
\caption{Operating space with an \supred{} edge in terms of (a) toroidal current (\Ip{}) and toroidal magnetic field (\Bt{}), (b) major radius (\Rnot{}) and minor radius (\aminor{}), (c) elongation (\elong{}) and average triangularity (\triavg{}). \supred{} operating points are in color, while an ELMing database is shown as black dots \cite{Thomsen2002}.}
\label{fig:basics}
\end{figure*}


\subsection{Basic Machine and Shaping Parameters}
\label{sec:eng}

The operating space in terms of toroidal current (\Ip{}) and toroidal field (\Bt{}) is shown in Fig.~\ref{fig:basics}(a). When considering all devices, a fairly wide range of values in both of these parameters is found. However, individual devices see more limited access to \supr{}. \diiid{} data reveals a relatively fixed ratio of \Ip{} to \Bt{}, indicative of a narrow range of safety factor (\qnf{}, from $\approx$ 3-4) owing to well known resonant window requirements \cite{Evans2008,Suttrop2018,Hu2020b}. \aug{} thus far has accessed the regime in only a limited range of \Ip{} and \Bt{}, which is partially related to the constraints imposed by central electron heating for high-Z impurity exhaust \cite{Angioni2017a}. \east{} is able to access \supr{} over a range of \Bt{}, but tends to favor operation in a more limited range of \Ip{}. \kstar{} has explored the most decoupled ranges of \Ip{} and \Bt{} \cite{In2019}.

Unlike the fairly wide variability in \Ip{} and \Bt{}, Fig.~\ref{fig:basics}(b) illustrates the degree to which geometric size of all devices accessing \supr{} is rather similar.  \devices{} are all mid-scale tokamaks with conventional aspect ratio, and as illustrated occupy a narrow range compared to the worldwide database. As will be shown, this fact is a key limitation of worldwide \supr{} research and strongly motivates the inclusion of RMP coils into future devices of different size, such as ITER \cite{Schaffer2008,Neumeyer2011,Daly2013} and JT-60SA \cite{Matsunaga2015,Yoshida2022}. When considering the regression analysis presented in Sec.~\ref{sec:scaling}, no size scaling will be possible. The inset in Fig.~\ref{fig:basics}(b) does show some modest variation in aspect ratio, which will be shown later to lead to variations in \In{} at constant \qnf{}.

A more varied picture is found in terms of plasma shaping, which covers a wide range due to variations both across as well as within machines. While \supr{} is accessed over a wide range of shapes, individual devices find strong shaping thresholds \cite{PazSoldan2019, Jeon2018IAEA}. \diiid{} has reported a \triavg{} threshold above which no \supr{} is found \cite{Gu2022}, while a lower \triavg{} threshold below which no suppression is found is reported in \aug{} \cite{Nazikian2016IAEA,Suttrop2018}. In \east{} and \kstar{}, progressing towards double-null divertor shapes also raises \triavg{}, with an upper limit in \triavg{} near DND shapes reported \cite{Gu2023}. It is at present unclear if \triavg{} plays a role in this threshold or if this physics might be responsible for the difficulties seen in spherical tokamaks. Thus far, no reports of thresholds in \elong{} have appeared in the literature. Due to its shape flexibility, \diiid{} has reported \supr{} in plasmas closely matching ITER's \elong{}, \triavg{}, \In{}, and \qnf{} \cite{Wade2015}.


\subsection{RMP Coil Configurations and Safety Factor Windows}
\label{sec:RMP}

\begin{figure*}
\begin{subfigure}{1\textwidth}
\begin{subfigure}{0.325\textwidth}
\centering
\includegraphics[width=1\textwidth]{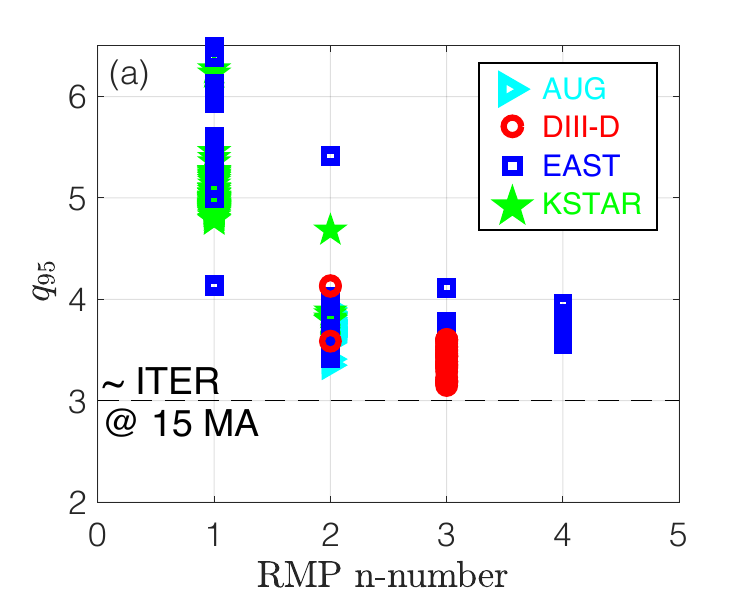}
\end{subfigure}
\begin{subfigure}{0.325\textwidth}
\centering
\includegraphics[width=1\textwidth]{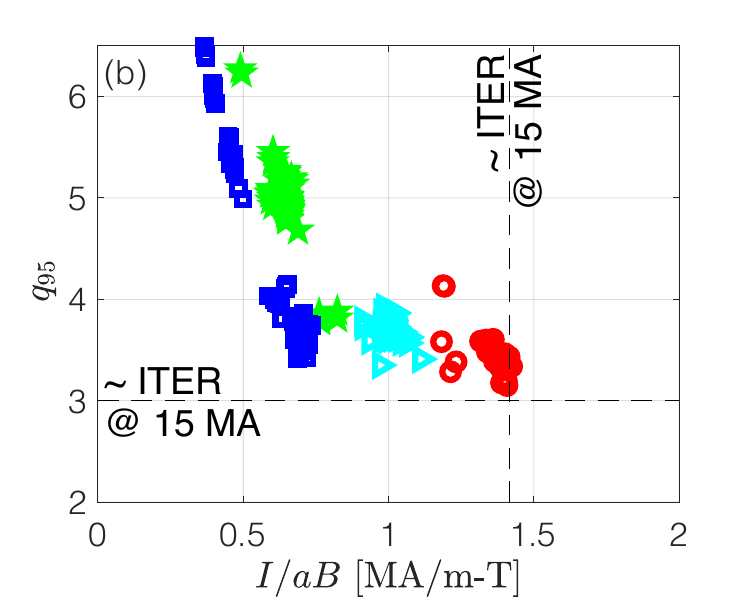}
\end{subfigure}
\begin{subfigure}{0.325\textwidth}
\centering
\includegraphics[width=1\textwidth]{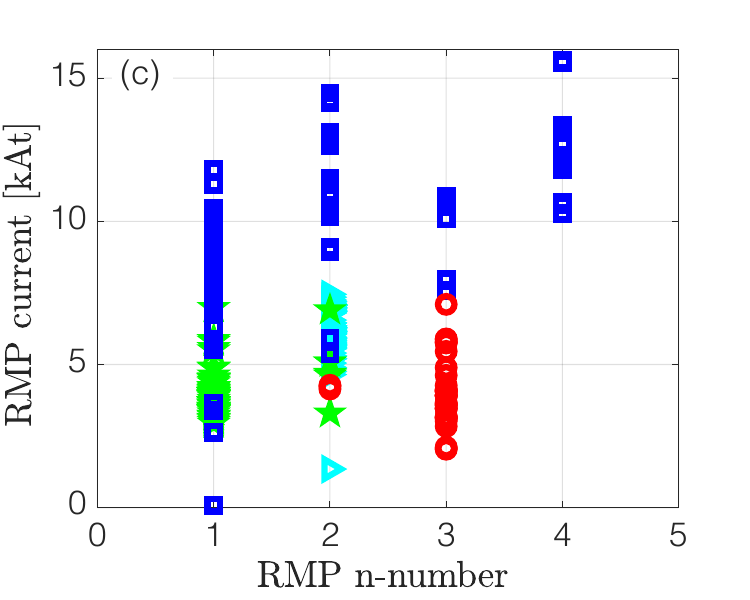}
\end{subfigure}
\end{subfigure}
 \vspace{-5 pt}
\caption{\supr{} window dependence on (a) RMP toroidal mode `n-number' and edge safety factor (\qnf{}), (b)  normalized current (\In{}) and \qnf{}, and (c) RMP n-number and RMP coil current. Different devices can use different n-numbers based on the number of coils per toroidal row.}
\label{fig:rmpcoil}
\end{figure*}

Each of the different experiments also feature different geometries of RMP coils, shown already in poloidal cross-section in Fig.~\ref{fig:devices}.  Both \aug{} and \east{} are equipped with two toroidal rows of 8 RMP coils each, above and below the midplane, favoring $n = 2, 4$, where $n$ is the toroidal mode number of the applied RMP field. \diiid{} is equipped with two similar rows, but with 6 coils each (favoring $n = 3$). \kstar{} has three toroidal rows, each with 4 coils, thus favoring $n = 1,2$ \cite{Kim2017a}. As can be seen in Fig.~\ref{fig:rmpcoil}(a), access to \supr{} differs in n-number and \qnf{} across devices. Access with $n = 2, 3$ \cite{Lanctot2013,Nazikian2015,Gu2019} has been observed in DIII-D, but not $n = 1$. \aug{} has thus far accessed \supr{} with only $n=2$. \kstar{}'s RMP coils are capable of both $n = 1, 2$, and both have been used to access \supr{}. The \east{} RMP coils are capable of $n = 1, 2, 3, 4$, and all have been used to access RMP-ELM suppression. As can be seen, low-n RMPs favor \supr{} at high \qnf{}, while high-n favors low \qnf{}. This is because the coupling to the resonant surfaces is poor for high-n RMPs at high \qnf{} \cite{Chen2022a}, while low-n RMPs risk driving disruptive $m/n = 2/1$ tearing instabilities at low \qnf{}, requiring careful tuning of the poloidal spectrum \cite{Park2018,Yang2020}. Note variations in the poloidal spectrum are not here further discussed, as the spectra to achieve the most favorable results is specific to each device's RMP coil geometry. Results further depend on whether the goal is simply maximizing the coupling to edge resonant surfaces, or reducing core surface coupling whilst maintaining sufficient edge coupling \cite{PazSoldan2015,Park2017,Zhou2018,Park2018,Ryan2018,Logan2021}.

Figs.~\ref{fig:rmpcoil}(a,b) both show clear windows in \qnf{}, inside of which \supr{} is possible. The window behavior in \qnf{} is a classic signature of the \supr{} phenomenon \cite{Evans2005,Evans2008a}, which has been recovered by simulations considering the alignment of the edge rational surface with the pedestal-top region \cite{Hu2021,Fitzpatrick2021}. Each device has gaps in \qnf{} where \supr{} is not achieved, though when considering all devices a wide range of \qnf{} values over which \supr{} is observed. Both \diiid{} and \aug{} have not yet accessed full \supr{} at high \qnf{}, though \diiid{} achieved type-I \supr{} at \qnf{} = 7 \cite{Fenstermacher2008} with $n=3$ fields (not shown). In the range of \qnf{} 4.8-6.2 is also where \aug{} first discovered strong ELM mitigation \cite{Suttrop2011}. Note very recent \east{} data, taken after the compilation of this database, has extended $n=4$ \supr{} to \qnf{} $\approx$ 3 \cite{Sun2023} and \qnf{} $>$ 4 \cite{Xie2023}. Also interesting is the generic accessibility of a window between \qnf{}$ = 3 - 4$. From the range of \In{} presented it can be surmised that \qnf{} is an appropriate characterization of the operating window, since the window does not depend strongly on \In{} (ie, it is invariant with aspect ratio).

It is also illustrating to consider the RMP current (in absolute kAt units) required to access \supr{} in these devices. Naturally, the coilsets are not equivalent in terms of their propensity to create the 3D topology necessary for ELM suppression. This is consistent with a very wide spread of RMP coil currents, and motivates a more physics-informed treatment of the RMP coil amplitude and coupling effects in the future.


\subsection{Pedestal Parameters and Density Limits}
\label{sec:corr}

\begin{figure*} 
\begin{subfigure}{1\textwidth}
\begin{subfigure}{0.4\textwidth}
\centering
\includegraphics[width=1\textwidth]{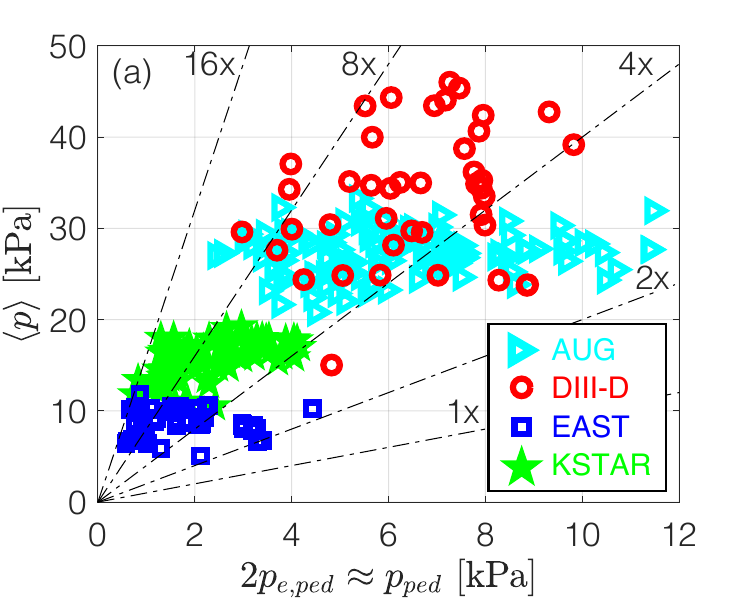}
\end{subfigure}
\begin{subfigure}{0.4\textwidth}
\centering
\includegraphics[width=1\textwidth]{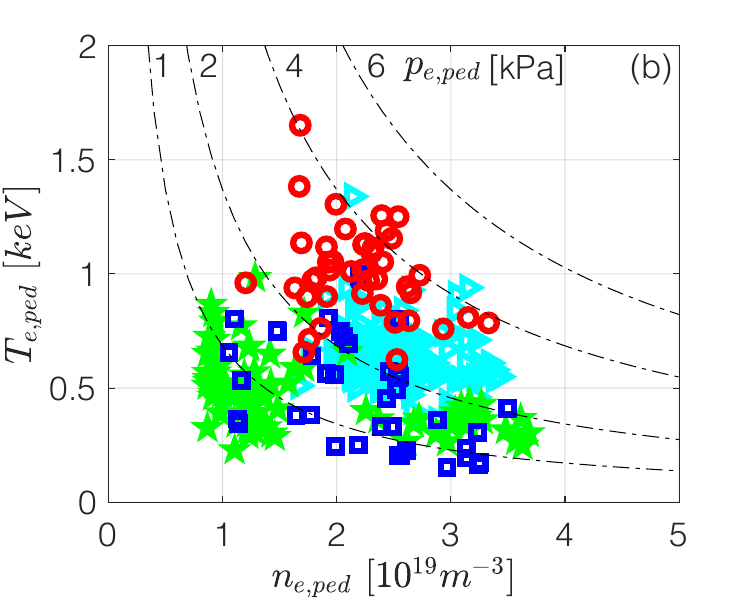}
\end{subfigure}
\begin{subfigure}{0.4\textwidth}
\centering
\includegraphics[width=1\textwidth]{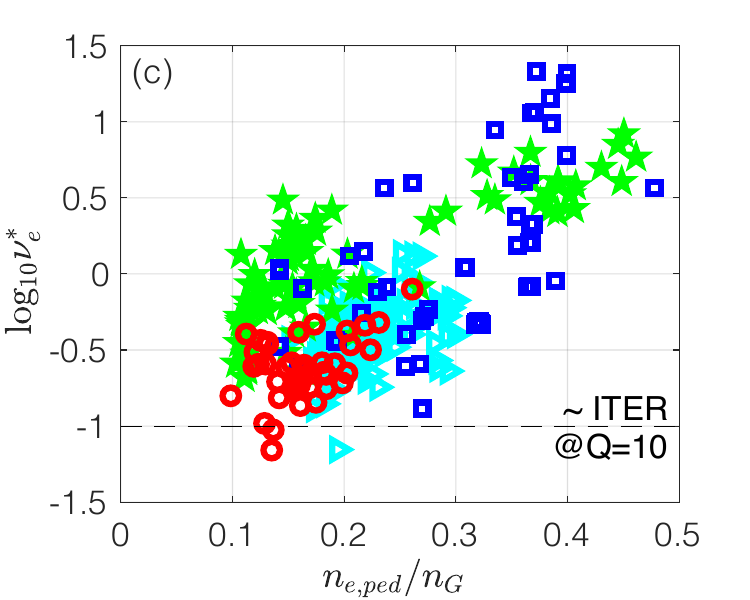}
\end{subfigure}
\begin{subfigure}{0.4\textwidth}
\centering
\includegraphics[width=1\textwidth]{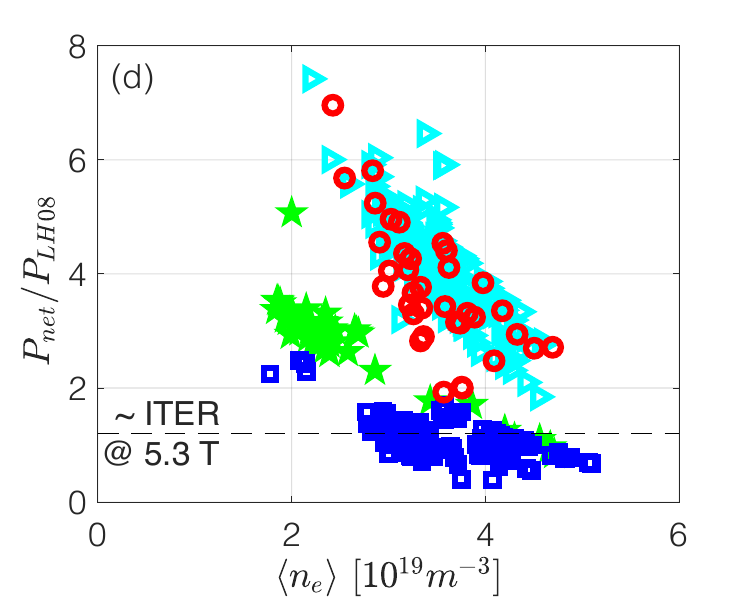}
\end{subfigure}
\end{subfigure}
 \vspace{-5 pt}
\caption{Pedestal parameters during \supr{}. (a) A wide range of pedestal pressure (\pped) with relatively invariant average pressure \pres{}. (b) Operating space of pedestal electron density (\neped{}) and temperature (\teped{}). (c) Pedestal-top electron collisionality (\nustar{}) and \neped{}, and (d) exhaust power divided by L-H threshold power (\PoverPLH{}) plotted against line-average density (\navg{}).}
\label{fig:ped}
\end{figure*}

Comparing pedestal parameters reveals consistent density limits and a wide range of other parameters. As a prelude, Fig.~\ref{fig:ped}(a) compares the pedestal pressure \pped{} to the volume-average pressure (\pres{}). Note \pped{} is taken to be 2\peped{} as measured by Thomson scattering or ECE, which is more experimentally reliable and is expected to be accurate in reactor conditions. A surprising result is that no significant correlation is seen between \pped{} and \pres{} in each device, with a relatively wide range of \pped{} not significantly levering \pres{}. This is contrary to the expectation that maximizing the pedestal optimizes the global performance \cite{Wagner2010}, and indicates distinct core confinement physics plays an important role in the global performance. These topics will be revisited in Sec.~\ref{sec:perf}.

A closer look at the electron pressure is shown in Fig.~\ref{fig:ped}(b), by comparing the contribution from the pedestal-top electron density (\neped{}) and temperature (\teped{}). The lower limit of these parameters is set by the L-H or H-L transition. A key finding arising from this work is that there is a consistent upper \neped{} limit found across all devices, which is particularly remarkable given the ranges of other parameters existing in the multi-device dataset. This will be discussed further in Sec.~\ref{sec:disc}. Another interesting finding is that the \pres{} ordering seen in Fig.~\ref{fig:ped}(a) is recovered in \teped{}. Each device, however, spans a range of \neped{}, and as expected \teped{} can be higher when \neped{} is lower.

One of the classic representations of the \supr{} operating space is in \neped{} and pedestal-top electron collisionality \cite{Kirk2015} (where \nustar{} is computed using the equation found in Ref. \cite{Sauter1999}), reproduced in Fig.~\ref{fig:ped}(c). Recall type-I ELM suppression can also be found at high \neped{} and \nustar{} in \aug{}\cite{Suttrop2011} and \diiid{}\cite{Moyer2005} (though small ELMs remain), but this data is not included in this database, which looks at suppression of all ELMs. An interesting result is the wide access in \nustar{} found in \east{} and \kstar{}, indicating perhaps a continuous transition between the low \nustar{} and high \nustar{} regime in those devices. Note the relative scarcity of intermediate \neped{} points in \kstar{} is thought to be due to sampling bias as opposed to a difficult to access region.

A final pedestal factor explored is the relationship of the exhaust power (\Pnet{}) to the H-mode power threshold \PoverPLH{}, where the LH08 scaling is taken from Ref. \cite{Martin2008}. In Fig.~ \ref{fig:ped}(d) this is plotted against the average density \navg{}.  It can be seen that \diiid{} and \aug{} operate at overall higher \PoverPLH{}, and that a decreasing trend is seen as \navg{} is rising. This is in part due to the inverse dependence of \navg{} in the LH08 power threshold scaling (\Pnet{} itself does not fall with \navg{}), but it does illustrate that the \supr{} phenomenon appears to exist at the approximate location that the low density branch of the H-mode threshold would be expected to exist \cite{Ryter2014}. Since the database does not contain corresponding L-mode points, it cannot be used to identify the L-H transition for these conditions.


\subsection{Limits to Electron Heating}
\label{sec:ech}

\begin{figure}
\begin{subfigure}{0.4\textwidth}
\begin{subfigure}{1\textwidth}
\centering
\includegraphics[width=1\textwidth]{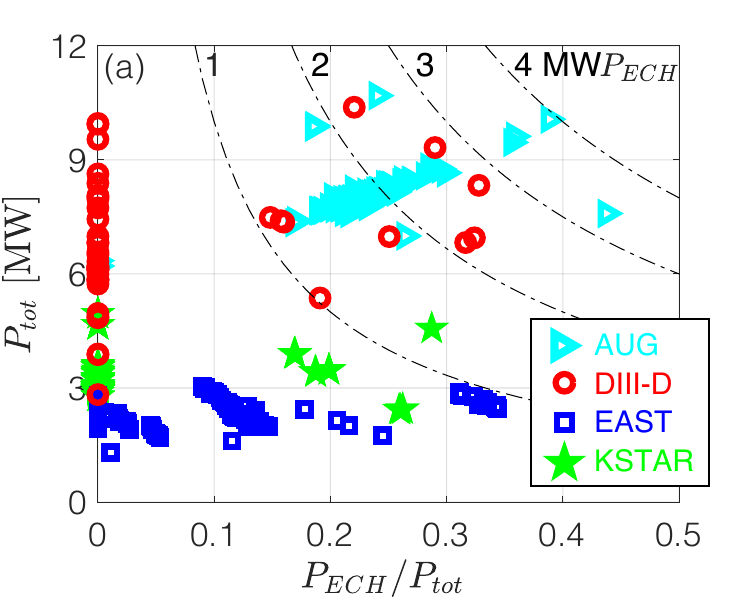}
\end{subfigure}
\begin{subfigure}{1\textwidth}
\centering
\includegraphics[width=1\textwidth]{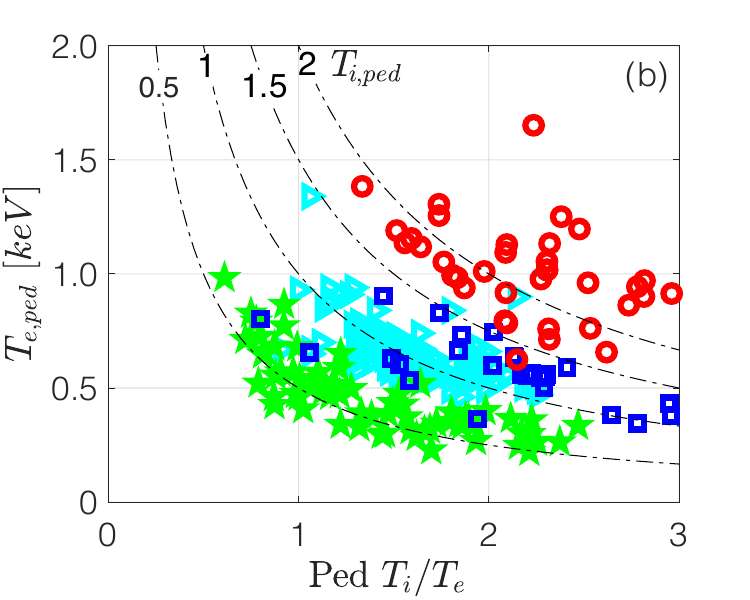}
\end{subfigure}
\end{subfigure}
\vspace{-5 pt}
\caption{Integration with electron heating: (a) operational space in terms of the ratio of ECH heating (\Pech{}) to total input power (\Ptot{}) plotted against \Ptot{} with dashed lines at constant \Pech{}. (b) Most data exists at high ion to electron temperature ($T_i/T_e$), owing to a relative lack of \Pech{}.}
\label{fig:ECH}
\end{figure}

Integration of \supr{} with electron cyclotron heated (ECH) plasmas is presented in Fig.~\ref{fig:ECH}. ECH will be the dominant power source for ITER pre-fusion power plasmas \cite{Loarte2021} as well as a reasonable substitute for bulk plasma heating from fusion-born alpha particles. This distinction is important because electron heating significantly modifies the turbulent transport properties of plasmas \cite{Petty1999,Prater2004,Kaye2013,Howard2016,Angioni2017,Grierson2018}. The importance of this physics is well appreciated, motivating worldwide studies of the impact of dominant electron heating on the \supr{} scenario, often limited by the availability of sufficient \Pech{}, as quantified in Fig.~\ref{fig:ECH}(a), where \PecoverPtot{} is always rather low.

On \aug{}, which has achieved the highest absolute \Pech{} and \PecoverPtot{}, there is evidence for a loss of \supr{} at the highest \PecoverPtot{} (around 0.5).  However, \diiid{} has not yet identified an operational limit within the range of \Pech{} injected \cite{Wade2015}. \east{} has similarly not found a limit with \Pech{}, and considering wave heating from lower hybrid and ion cyclotron heating, has accessed \supr{} in fully wave-heated discharges \cite{Sun2016}.  Experiments at \kstar{} are ongoing with findings still to be reported \cite{Lee2020a}. Considering the pedestal impact of \Pech{}, Fig.~\ref{fig:ECH}(b) shows a high proportion of the data exists at relatively high ion temperature ($T_i$) as compared to $T_e$, again indicating a preponderance of data with strong ion heating and potentially also significant main ion fuel dilution. Interestingly, \aug{} has previously reported an operational boundary when $T_i$ falls below $T_e$ \cite{Suttrop2018}, and the origin of this continues to be investigated.

\vspace{-10 pt}
\section{Plasma Performance with an RMP-ELM Suppressed Edge}
\label{sec:perf}

The plasma performance of discharges with \supred{} edges is surveyed, first in terms of normalized parameters such as confinement quality factors (\HH{}, \HL{}) and normalized pressure (\betan{}), followed by absolute performance metrics such as the average pressure (\pres{}) and confinement time (\taue{}). Due to the wide range of \Ip{}, \Bt{}, \density{}, and \Ptot{} available, scaling against these parameters is pursued to provide insight on the confinement extrapolation to future devices.


\begin{figure*}
\begin{subfigure}{1\textwidth}
\begin{subfigure}{0.325\textwidth}
\centering
\includegraphics[width=1\textwidth]{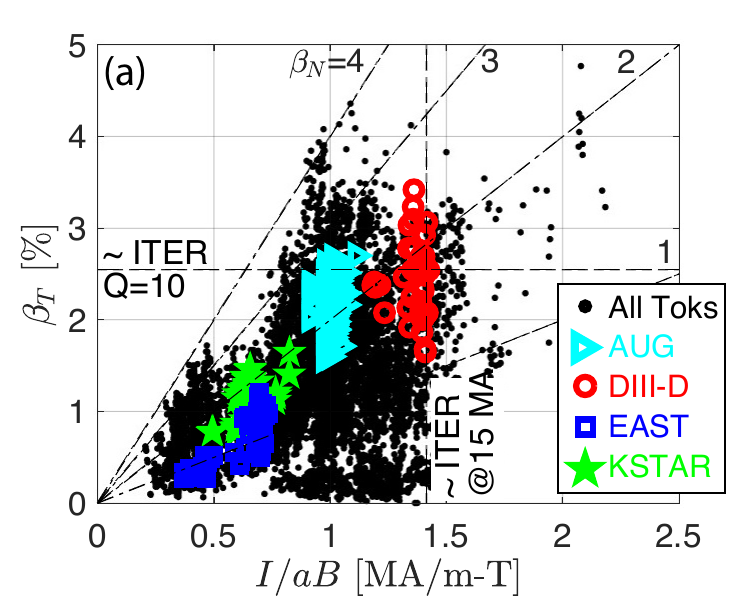}
\end{subfigure}
\begin{subfigure}{0.325\textwidth}
\centering
\includegraphics[width=1\textwidth]{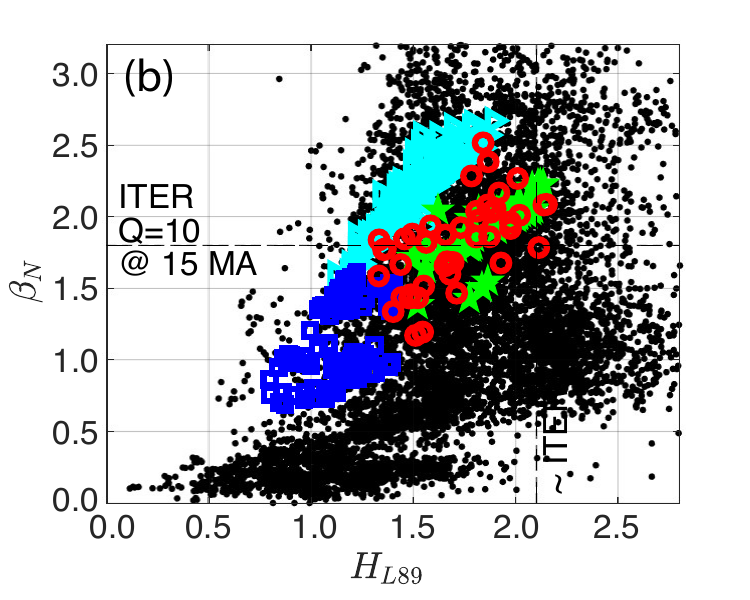}
\end{subfigure}
\begin{subfigure}{0.325\textwidth}
\centering
\includegraphics[width=1\textwidth]{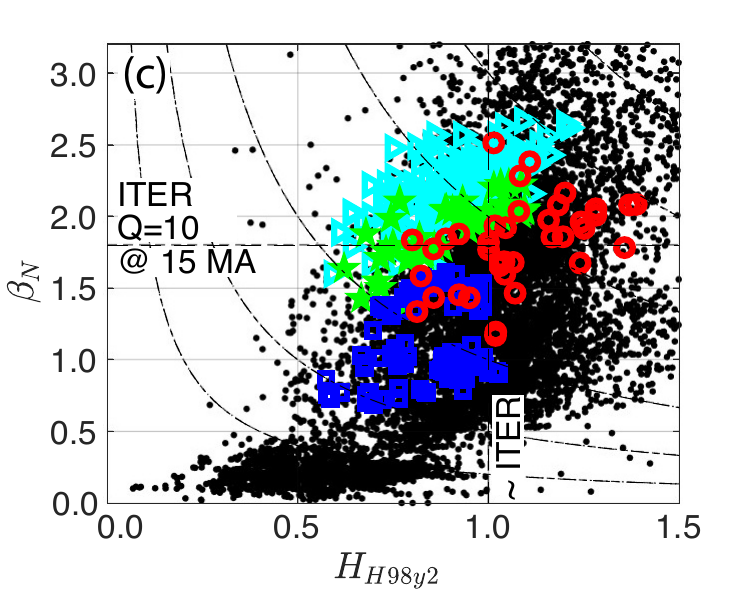}
\end{subfigure}
\end{subfigure}
 \vspace{-5 pt}
\caption{Normalized plasma performance in terms of (a) Troyon core stability given by toroidal beta (\betat{}) and normalized current (\In{}) with dash-dot lines indicating constant normalized pressure (\betan{}). Confinement quality factors using (b) L-mode scaling laws (\HL{}) or (c) H-mode scaling laws (\HH{}) as a function \betan{}. Dashed lines indicate ITER 15 MA Q=10 targets. Data from the global tokamak database is also included.}
\label{fig:Hfactors}
\end{figure*}

\subsection{Normalized Performance}
\label{sec:norm}

The core stability of tokamak plasmas is classically represented by the Troyon stability diagram, \cite{Troyon1984}, shown in Fig.~\ref{fig:Hfactors}(a). These diagrams plot the toroidal beta (\betat{}) against the normalized current (\In{}) with diagonal dashed lines representing the normalized beta (\betan{}). The \supred{} data occupies a region of the Troyon diagram at intermediate \betan{}, with lower values presumably falling into L-mode, and higher values limited by stability or confinement. Across all devices a wide range of \In{} is found, as was also shown in Fig.~\ref{fig:rmpcoil}(b). Overall the normalized stability of \supred{} plasmas in terms of these parameters is consistent with the requirements of ITER as indicated by the dashed lines.

Normalized plasma performance is usually described in terms of the confinement quality `H-factor'. Fig.~\ref{fig:Hfactors}(b) highlights the normalized pressure (\betan{}) against the H-factor derived for L-mode plasmas (\HL{}) \cite{Yushmanov1990}. The \HL{} data highlights the typically seen intermediate level of confinement between L-mode (\HL{} $\approx 1$) and H-mode (\HL{} $\approx 2$). The \HL{} factors also span a wide range in any individual device, that is overall between $\approx$ 1 and 2 for all devices. A strong correlation of \betan{} and \HL{} is seen in each device, indicating power degradation is not as severe as expected by the \HL{} law. Section \ref{sec:scaling} will use this data to explore the validity of the \HL{} scaling law for these plasmas and to discuss extrapolation to next-step devices. A notable feature is the presence of two clusters of \east{} data, with the lower \betan{} cluster being the lower power dominantly wave-heated discharges, motivating additional studies on the optimization of the \supr{} scenario in these conditions. Very recently, and after the compilation of this database, higher \betan{} values (2.65) have been achieved in \kstar{}\cite{Kim2023a}. The \betan{} = 1.8 target of ITER is well within the multi-device dataset and is approximately accessed by all devices.

Comparison to the H-mode `H98y2' scaling law \cite{Doyle2007} is presented in Fig.~\ref{fig:Hfactors}(c). Similar trends in confinement quality between \HH{} and \HL{} are observed, and again the requisite values for ITER are well within the multi-device database. Section \ref{sec:scaling} will explore extrapolation based on the observed \taueth{} and H-mode expectations. Note presenting \HH{} values requires computing the fast ion fraction, which will be shown in Sec.~\ref{sec:scaling} to be a key uncertainty. For \diiid{} and \aug{}, automated between-shot transport routines \cite{Heidbrink1994,Pankin2004} extract the fast ion pressure and are expected to be fairly robust. For \east{}, which is dominated by radio-frequency heating, a fast-ion fraction of 0\% is assumed. For \kstar{}, the fast ion fraction is taken to be uniformly 40\%, in line with previous transport calculations \cite{Kim2023,Na2020}.



\begin{figure*}
\begin{subfigure}{1\textwidth}
\begin{subfigure}{0.325\textwidth}
\centering
\includegraphics[width=1\textwidth]{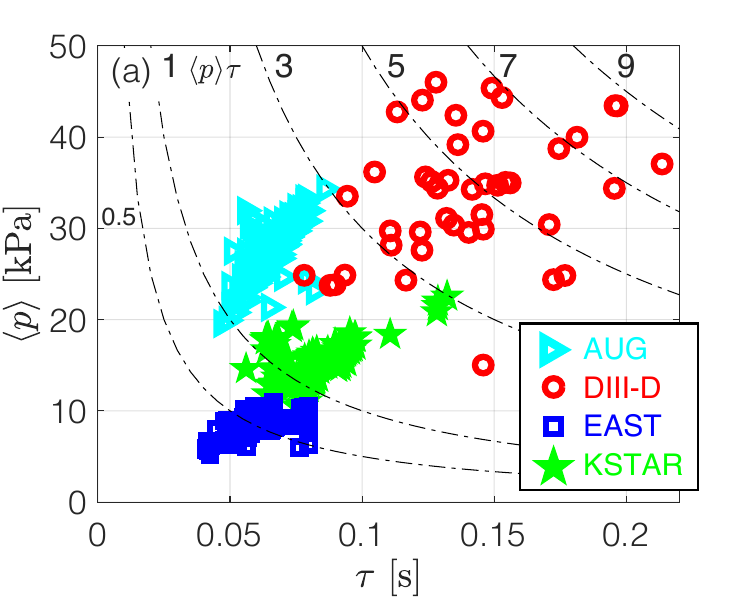}
\end{subfigure}
\begin{subfigure}{0.325\textwidth}
\centering
\includegraphics[width=1\textwidth]{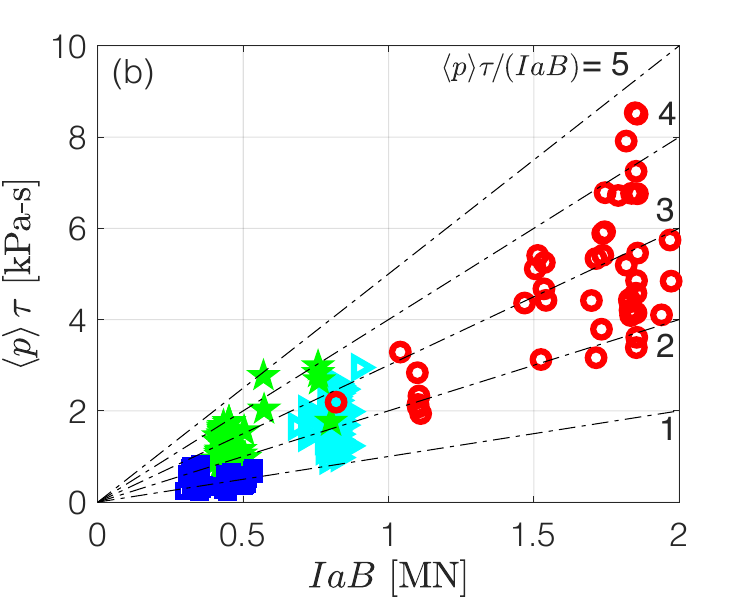}
\end{subfigure}
\begin{subfigure}{0.325\textwidth}
\centering
\includegraphics[width=1\textwidth]{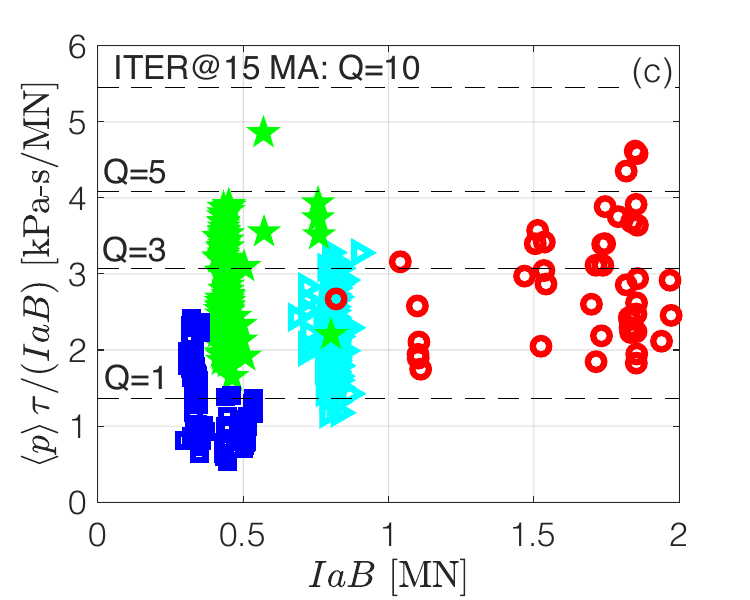}
\end{subfigure}
\end{subfigure}
 \vspace{-5 pt}
\caption{Absolute plasma performance represented by: (a) average pressure (\pres{}) and confinement time (\taue{}), with dash-dot lines at constant triple product (\lawson{}).  (b) Triple product \lawson{} referenced to \IaB{}, showing an increasing trend. (c) \lawson{} normalized to \IaB{} referenced to \IaB{}. Dashed lines reference fusion gain Q values at 15 MA in ITER.}
\label{fig:abs}
\end{figure*}

\subsection{Absolute Performance}
\label{sec:abs}

While normalized plasma performance naturally compares the data to the expectations of plasma stability and confinement quality, this approach can also mask underlying trends. Discussion of the absolute plasma performance offers additional insight. The most straight forward way to measure plasma performance is Lawson's triple product \cite{Lawson1957}, here taken as \lawson{}, which substitutes the average pressure for the more direct core ion temperature and fuel density. \lawson{} is more robust across the multi-device dataset, and further removes some bias towards highly peaked profiles, which may be more reactive in the core, but would produce less fusion power overall.

The triple product \lawson{} can be separated into its constituent \pres{} and \taue{}, as shown in Fig.~\ref{fig:abs}(a). An interesting feature is the different relative mixes of \pres{} and \taue{} across devices, and in particular all devices except \diiid{} occupy predominantly similar \taue{} ranges. Also interesting is the clustering of data from each device with relatively little overlap. This highlights the important role of the operational window in the observed performance, as will be further elaborated in Sec.~\ref{sec:disc}. Considering the highest \lawson{} discharges in \diiid{}, they are achieved by exploiting hysteresis in the bifurcation into RMP-ELM suppression, reducing the RMP current after ELM suppression is accessed \cite{Laggner2020}. Note the \lawson{} required for net energy gain and ignition is several atmosphere-seconds, varying based on the ion temperature \cite{Wurzel2022}.

A valuable normalization metric for \lawson{} is found to be \IaB{}, the product of the plasma current (\Ip{}), minor radius ($a$), and toroidal field (\Bt{}) and which has the units of force [MN]\cite{Snyder2019}. As shown in Fig.~\ref{fig:basics}, $a$ is similar across devices, while \Bt{} varies over about a factor of 2, and \Ip{} especially varying over a factor of 4, yielding an overall 7x variation in \IaB{}. This normalization metric also highlights the rather large extrapolation to ITER (\IaB{} $\approx 160$), and SPARC (\IaB{} $\approx 60$). Perhaps surprisingly, as shown in Fig.~\ref{fig:abs}(b) the peak \lawson{} is well-captured by the \IaB{} normalization across the multi-device \supr{} database, which is to say the peak \lawson{} values are found to increase with \IaB{}. Further discussion of the \snyder{} metric is found in Refs \cite{Snyder2019,PazSoldan2021}.

Figure \ref{fig:abs}(c) shows the absolute fusion performance metric \lawson{} normalized against the \IaB{} metric. Most data is found to occupy values of \snyder{} between 2 and 4. As with the \betan{} data in Fig.~\ref{fig:Hfactors}(b), there is a separate cluster of data from \east{} which is distinct due to the dominant wave heating of those points, compared to the dominant NBI heating of the data from \diiid{}, \aug{}, and \kstar{}. Further, since the \lawson{} target for ITER operating at 15 MA is known\cite{Lehnen2015} (\taueth{} $\approx$ 3 s, stored energy $\approx$ 350 MJ, volume $\approx$ 840 m$^{-3}$), as is the \IaB{}, the \snyder{} parameter can be directly related to an expectation of the fusion gain, Q, which is also included in Fig.~\ref{fig:abs}(c). This finds the peak fusion performance shy of the Q=10 target, with the preponderance of data existing between Q=1 and Q=5. These projections are rather crude however, and should not be taken beyond the high-level conclusion that the absolute performance is in the range of what is needed for future devices such as ITER.


\begin{figure}
\centering
\includegraphics[width=0.49\textwidth]{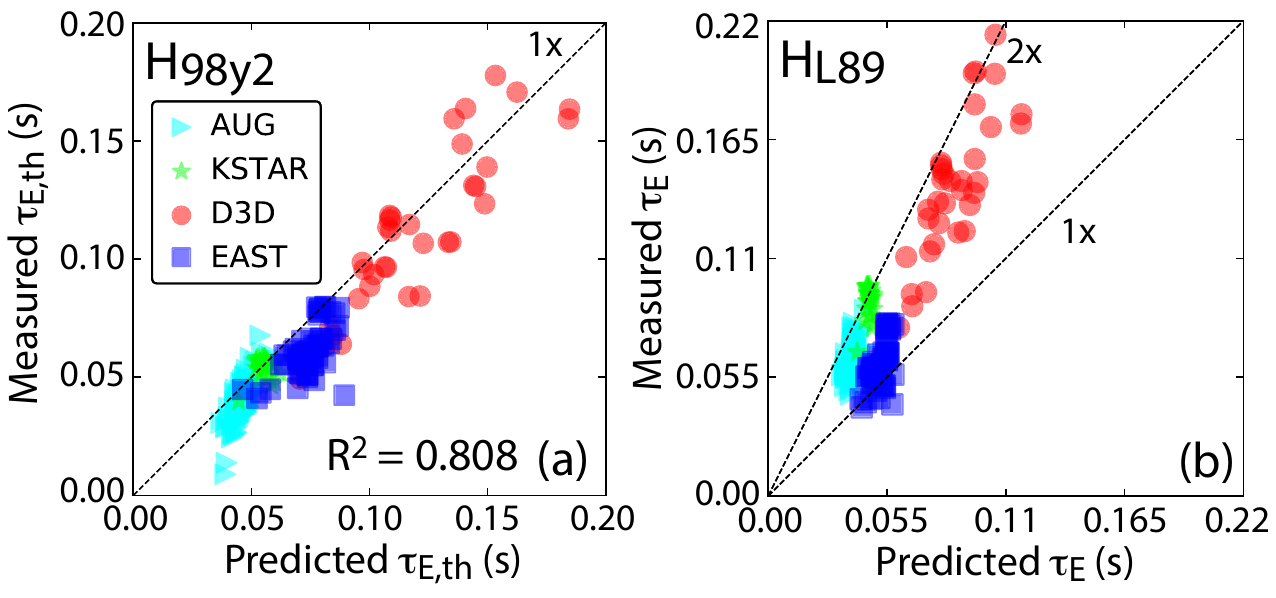}
\vspace{-20 pt}
\caption{Comparison of measured \taueth{} and \taue{} to scaling law predictions for (a) H-mode and (b) L-mode.}
\label{fig:scale1}
\end{figure}

\subsection{Confinement Scaling}
\label{sec:scaling}

Comparison of the observed confinement in \supred{} discharges is made to the well-established existing confinement scaling laws, and an assessment of improved scaling laws derived from the assembled databases is presented. Figure \ref{fig:scale1} presents the consistency of the \supred{} experimentally measured confinement time with the predictions from the established H-mode \HH{} scaling law \cite{Doyle2007,IPBch2} and \HL{} scaling law \cite{Yushmanov1990}. The \HH{} scaling requires estimating the fast ion contribution to the stored energy. As discussed in Sec.~\ref{sec:norm}, the certainty in the fast ion fraction varies across the multi-device dataset, and the approximations for \east{} and \kstar{} taken in Sec.~\ref{sec:norm} are retained. Within these uncertainties, a reasonable agreement is found with an R$^2$ value of 0.808 obtained. Turning to the \HL{} scaling law \cite{Yushmanov1990}, as was also shown in Fig.~\ref{fig:Hfactors}(b), the data is found to be intermediate between \HL{} of 1 and 2 as indicated in Fig.~\ref{fig:scale1}(b). An increasing slope is found, with the larger absolute confinement times trending towards \HL{} of 2. These observed differences in slope motivate exploring new scaling laws via engineering parameter regression for \supred{} plasmas \cite{Verdoolaege2021}. Note uncertainty in the fast-ion fraction does not impact projections based on \taue{} and \HL{}. 

\begin{figure} 
\centering
\includegraphics[width=0.49\textwidth]{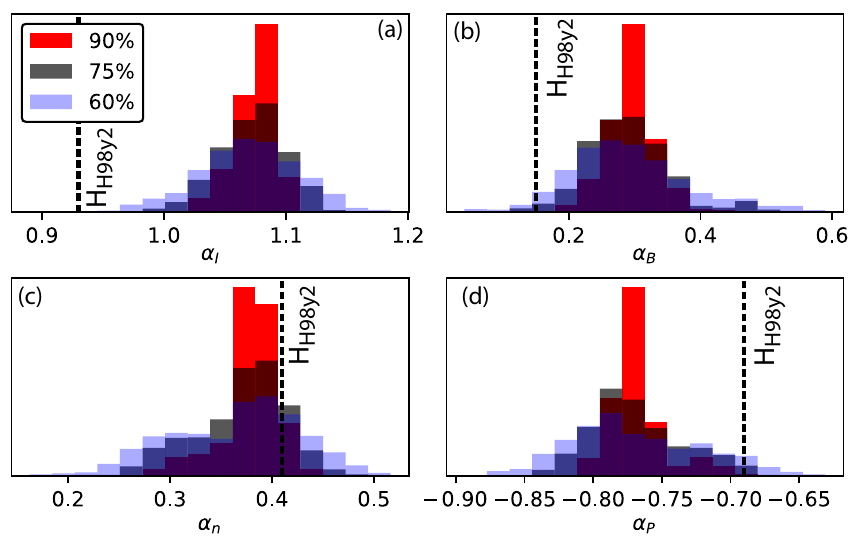}
 \vspace{-20 pt}
\caption{Probability density function (PDF) of the scaling law exponents for \taueth{} in terms of (a) \Ip{}, (b) \Bt{}, (c) \nebar{} and (d) \Paux{}, upon randomly sampling 90\%, 75\%, and 60\% of the full database. Dashed lines show expecatation from \HH{}.}
\label{fig:scaleHH}
\end{figure}

Exponential scaling coefficients are derived by deploying linear least-squares fitting of the confinement times with fast ion correction (\taueth{}) and without (\taue{}) to the logarithm of the engineering parameters of plasma current (\Ip{}), toroidal field (\Bt{}), average density (\nebar{}), and auxiliary heating power (\Paux{}). To alleviate sampling bias present in the database, kernel density estimation is used to assign weights to data points such that poorly-sampled regions of the parameter space are weighted more strongly than highly-sampled regions \cite{Logan2020a}. Using this technique, exponential scaling coefficients $\alpha_I$ for \Ip{}, $\alpha_B$ for \Bt{}, $\alpha_n$ for \nebar{}, and $\alpha_P$ for \Paux{} can be obtained. As shown in Fig.~\ref{fig:basics}(b), there is minimal size variation across devices, thus no attempt is made to include geometry fitting terms. Further, as discussed in Ref. \cite{Verdoolaege2021}, engineering parameter regression underestimates the uncertainties involved, which arise not from the statistical goodness-of-fit, but rather from the completeness of the underlying databases. To present the confinement scaling results along with a realistic assessment of uncertainty, the fitted exponential scaling coefficients are presented statistically across many different regression attempts on randomly selected 90\%, 75\%, and 60\% subsets of the full database prepared for this study. The full dataset yields a modestly narrower distribution than the 90\% subset.

Results are shown in Fig.~\ref{fig:scaleHH} for \taueth{} scaling along with the expectation from \HH{}. Each randomized subset regression yields its own scaling exponents, which can vary significantly, and are thus presented as probability density functions.  Notwithstanding the observed variability, some aspects appear robust in the present dataset: an approximately linear \Ip{} scaling is revealed, along with an \nebar{} scaling consistent with \HH{}, and a slightly stronger \Bt{} dependence and more severe power degradation than expected from \HH{} scaling.

Note that the scaling of \taueth{} introduces uncertainty from the magnitude of the fast ion contribution, which is subtracted from the stored energy to arrive at \taueth{}. The sensitivity to the magnitude of the fast ion correction was explored in dedicated scans (not shown). For \kstar{}, decreasing the fast ion fraction resulted in reduced $R^2$ values and a weaker exponent on fitted \nebar{}. This sensitivity is due to \kstar{} containing the lowest \nebar{} points in the entire multi-device database. For \east{}, the $R^2$ value monotonically increased as the fast ion fraction decreased, supporting the choice of using no fast ions for the regression analysis. The power degradation and density exponent also increased as the \east{} fast ion fraction decreased. Overall, the fast ion fraction is identified as a dominant uncertainty for the multi-device \taueth{} and \HH{} scaling exercises.

\begin{figure} 
\centering
\includegraphics[width=0.49\textwidth]{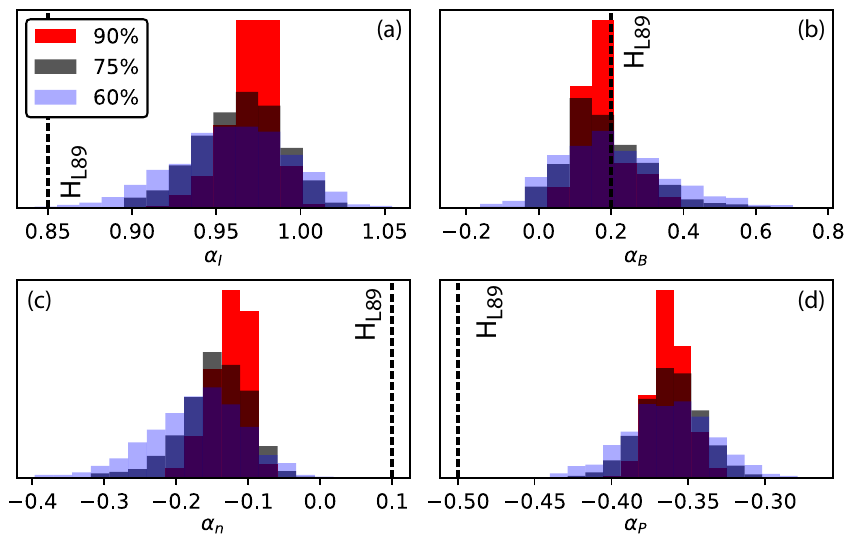}
 \vspace{-20 pt}
\caption{PDF of the scaling law exponents for \taue{} in terms of (a) \Ip{}, (b) \Bt{}, (c) \nebar{} and (d) \Paux{}, upon randomly sampling 90\%, 75\%, and 60\% of the full database. Dashed lines show expectation from \HL{}.}
\label{fig:scaleHL}
\end{figure}

Scaling of \taue{} and comparison to the \HL{} scaling law avoids the uncertainty associated with determining the fast ion content. As shown in Fig.~\ref{fig:scaleHL}, some variation in scaling law exponent is found for \taue{} regression as compared to \taueth{} regression. The linear \Ip{} dependence is preserved. A significantly weaker power degradation is found for \taue{} as compared to \taueth{}, even in excess of the expectation from \HH{} to \HL{}. This suggests much of the heating power goes into the fast ion population, which is not surprising given these are low density predominantly NBI-heated plasmas. Most surprising is the exponent on \nebar{}, which robustly is found to be weakly negative, in contrast to the weakly positive expectation from \HL{}. The reason is not understood, though in absolute terms the difference from \HL{} is no worse than in the other parameters.

\begin{figure} 
\centering
\includegraphics[width=0.49\textwidth]{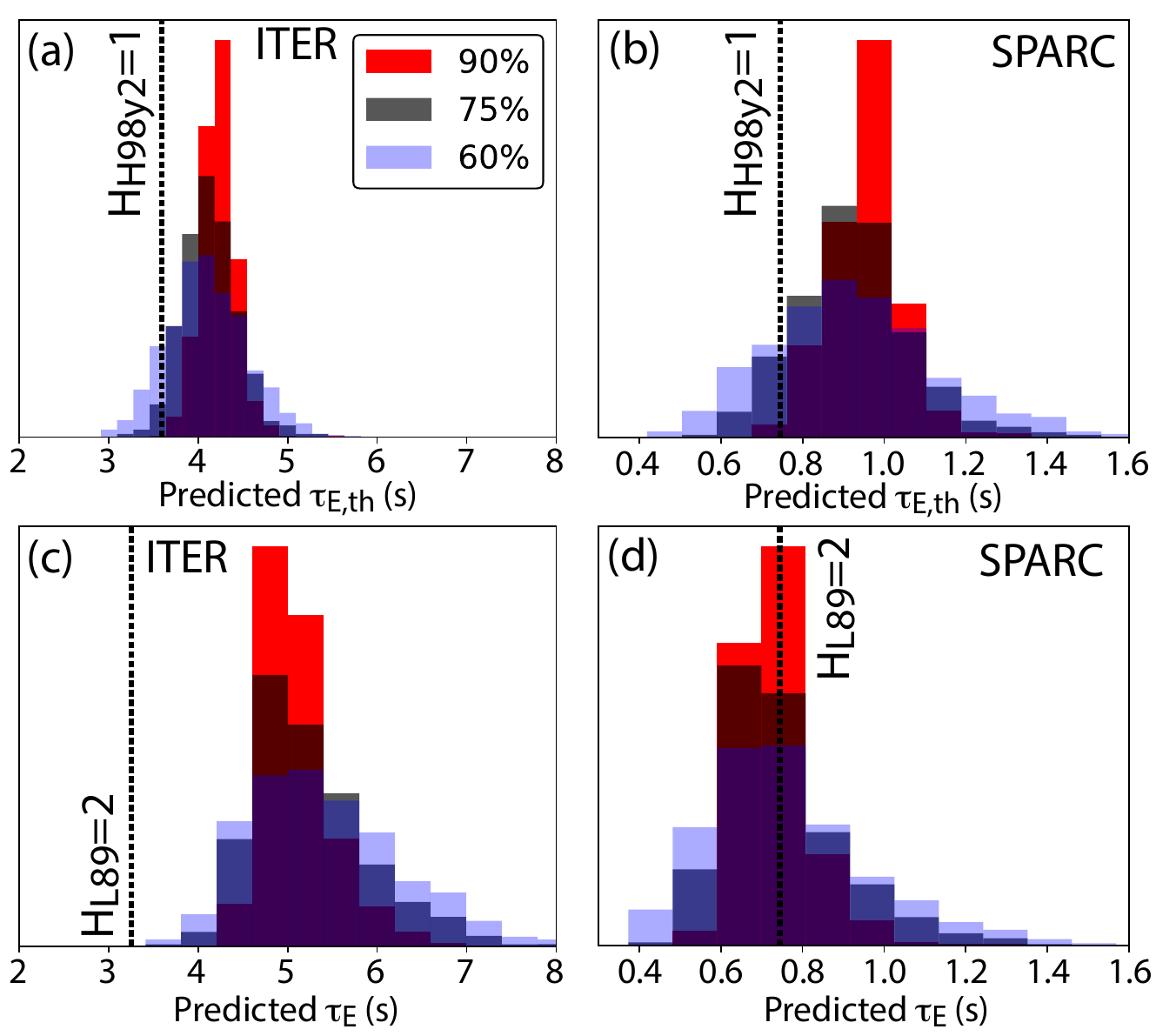}
 \vspace{-15 pt}
\caption{Extrapolation of \taueth{} (Fig.~\ref{fig:scaleHH}) and \taue{} scaling (Fig.~\ref{fig:scaleHL}) to ITER and SPARC parameters as compared to the existing \HH{}=1 and \HL{}=2 expectation, expressed as PDFs.}
\label{fig:ITER}
\end{figure}

With the same statistical approach used to describe the exponential scaling coefficients, the results of \taueth{} (Fig.~\ref{fig:scaleHH}) and \taue{} (Fig.~\ref{fig:scaleHL}) scalings are used to perform a projection to the upcoming tokamaks under construction, ITER \cite{Aymar2001} and SPARC \cite{Creely2020}, as shown in Fig.~\ref{fig:ITER}. In order to perform the extrapolation to ITER, physical size scaling exponents are taken from the pre-existing \HH{} or \HL{} scaling, introducing a significant uncertainty to those projections. Interestingly, this caveat is relatively unimportant for SPARC, since its physical size is comparable to the devices used in this regression. The standard isotope mass scaling is also used assuming a 50:50 deuterium tritium mix. For both devices, the \taueth{} extrapolation is modestly above the existing projections from \HH{} = 1, but well within the uncertainties characterized by the randomized subsets. Thus, the variations in exponents seen in Fig.~\ref{fig:scaleHH} are found to cancel out to some degree. Extrapolations based on \taue{} scaling for SPARC (Fig.~\ref{fig:ITER}[d]) are also found to be fully overlapping with expectations, indicating cancellation of the density and power exponent deviations for that device. For ITER however (Fig.~\ref{fig:ITER}[c]), a significantly higher confinement time is predicted, likely due to the higher input power and lower density expected for that device. Also worth noting is the relatively large uncertainty for these projections, indicating the extrapolation is relatively unconstrained. Gaps in the input data utilized for the scaling exercises will be discussed in the following section. Despite the uncertainty, it is clear that the inclusion of an \supred{} edge did not produce dramatically pessimistic confinement projections for either ITER or SPARC.

Finally, a similar analysis was also conducted for regression analyses including the effect of plasma rotation, which is also available for a subset of the multi-device RMP database. While for brevity the results are not shown in detail, the generic effect of including rotation was found to be to reduce the power degradation extracted from the dataset. Instead, a positive correlation of the confinement with rotation was observed, with exponents in the range of 0.1-0.2 found. The utilization of these regressions for extrapolation requires an accurate assessment of the rotation expected in ITER and SPARC while the edge is \supred{}, a question which is still under active study \cite{Chrystal2020,Rodriguez-Fernandez2020}.


\section{Discussion and Conclusions}
\label{sec:disc}


A remarkable finding of this effort is the consistency of the upper \neped{} limit observed across all devices. The consistency is observed despite a wide range of machine parameters, RMP coil geometries, and operational scenarios. Considering individual devices such as \diiid{}, differences in the \neped{} threshold have been reported as \teped{} is varied \cite{PazSoldan2019}, however this does not extend to the full multi-device database. Unless the finding is coincidental, the commonality across devices suggests an important role for the device size or \rhostar{} in this limit. Taking the simplifying rough approximation that the separatrix density is proportional to \neped{}, the invariance of the \neped{} threshold with \Ip{} greatly facilitates divertor integration at low \Ip{}. This is because the heat flux width scales inversely with \Ip{} \cite{Eich2013}, spreading the divertor power dissipation over a wider area at low \Ip{} (assuming fixed separatrix density). These dependencies provide an explanation for the relative difficulty of \aug{} and \diiid{} achieving edge-integrated \supr{}, as compared to \east{} and \kstar{} \cite{Petrie2011, Suttrop2018, Jia2021a}, though the dynamics of the seeded impurities also play an essential role in divertor integration \cite{Sun2021,Shin2023} as does the divertor geometry. Overall, additional effort is required to understand the \neped{} threshold, as its scaling to ITER and future devices is the key question confronting the possibility of integrating a dissipative divertor with the RMP scenario in high current scenarios.

Considering the confinement scaling exercise, one key finding shown in Fig.~\ref{fig:ped}(a) is the important role the core confinement (as opposed to the pedestal) is playing in the overall plasma performance of the \supred{} plasmas. This observation is thought to be related to the relatively low density of these plasmas, coupled with their neutral-beam injection (NBI) dominated heating mixes. NBI heating at low density tends to impart significant momentum to the plasma, bringing in additional dependencies to the confinement (such as core ExB shear \cite{Burrell1997,Burrell2020a} and hot ion modes \cite{Kotschenreuther1995,Keilhacker1999,Petty1999}) that are less likely to extrapolate to future devices. To address this issue, exploration of \supr{} under dominant electron heated low input torque regimes is an active area of study in the world program, as described in Sec.~\ref{sec:ech}.

Another general observation from this exploration is the limitations in what can be achieved with multi-device comparisons when the device size is relatively similar, as was shown in Fig.~\ref{fig:basics}(b). Engineering parameter regression cannot be extended to the major or minor radius, adding uncertainty to its use to extrapolate to ITER or other larger devices, though this is not an issue for SPARC. Installation of RMP coils on devices of larger major radius is essential for empirical extrapolation to ITER. This could be provided by JT-60SA in the coming years \cite{Ishida2011,Matsunaga2015}. Installation of RMP coils on a smaller high-field device, such as COMPASS-U \cite{Logan2021a,Vondracek2021}, would also improve the ability to conduct empirical extrapolations. 


\section*{Acknowledgments}

This work was the result of dedicated database extraction by the authors from \devices{}. As such, the experimental teams from each of \devices{} deserve explicit recognition for their efforts in producing the plasmas summarized here. Additionally, the authors thank Dr. A.O. Nelson for assistance with database curation.

{\small{
This material is based upon work supported by the U.S. Department of Energy, Office of Science, Office of Fusion Energy Sciences, using the DIII-D National Fusion Facility, a DOE Office of Science user facility, under Award(s) DE-FC02-04ER54698, DE-SC0021968, DE-SC0022270, and DE-SC0020298, DEAC02-09CH11466 (Princeton Plasma Physics Laboratory). Work also supported by the Chinese Academy of Sciences. This research was also supported by the R\&D Program of `KSTAR Experimental Collaboration and Fusion Plasmas Research (EN2301-14)' through the Korea Institute of Fusion Energy (KFE) funded by Government funds.  This work has been carried out within the framework of the EUROfusion Consortium, funded by the European Union via the Euratom Research and Training Programme (Grant Agreement No 101052200 — EUROfusion).

Disclaimer: This report was prepared as an account of work sponsored by an agency of the United States Government. Neither the United States Government nor any agency thereof, nor any of their employees, makes any warranty, express or implied, or assumes any legal liability or responsibility for the accuracy, completeness, or usefulness of any information, apparatus, product, or process disclosed, or represents that its use would not infringe privately owned rights. Reference herein to any specific commercial product, process, or service by trade name, trademark, manufacturer, or otherwise does not necessarily constitute or imply its endorsement, recommendation, or favoring by the United States Government or any agency thereof. The views and opinions of authors expressed herein do not necessarily state or reflect those of the United States Government or any agency thereof.  Views and opinions expressed are however those of the author(s) only and do not necessarily reflect those of the European Union or the European Commission. Neither the European Union nor the European Commission can be held responsible for them.}}


\bibliography{library}



\end{document}